\documentclass[ALICE,manyauthors]{cernphprep}
\newcommand{\jpsi}{\ensuremath{\mathrm{J}/\psi}}

\newcommand{\pb}{Pb-Pb}
\newcommand{\pt}{$p_{\rm T}$}
\newcommand{\snn}{$\sqrt{s_\mathrm{NN}}$}

\newcommand{\acceps}{\ensuremath{(\mathrm{Acc}\times\varepsilon)_{\mathrm{J}/\psi}}}
\newcommand{\minv}{$M_\mathrm{inv}$}
\usepackage{rotating}
\usepackage{lineno}
\usepackage{color}
\begin{document}%
%
%
\begin{titlepage}
\PHnumber{2012-270}                 
\PHdate{16 Sep 2012}              
%
%
\title{Coherent \jpsi~photoproduction in ultra-peripheral Pb--Pb collisions at \newline \snn = 2.76 TeV}
\ShortTitle{\jpsi~photoproduction in ultra-peripheral Pb--Pb collisions}   
%
\Collaboration{ALICE Collaboration%
         \thanks{See Appendix~\ref{app:collab} for the list of collaboration
                      members}}
\ShortAuthor{ALICE Collaboration}      
\begin{abstract}
The ALICE collaboration
has made the first measurement at the LHC of \jpsi~photoproduction in ultra-peripheral \pb-collisions 
at \snn= 2.76 TeV. The \jpsi~is identified via its dimuon decay in the
forward rapidity region with the muon spectrometer for events where the hadronic activity is
required to be minimal.
The analysis is based on an event sample corresponding to an integrated luminosity
of about 55 $\mu\rm{b}^{-1}$. The cross section for coherent
\jpsi~production in the rapidity interval -3.6 $<$ $y$ $<$ -2.6 is measured to be 
$\mathrm{d}\sigma_{J/\psi}^{\mathrm{coh}} /\mathrm{d}y = 1.00 \pm 0.18(\mathrm{stat}) ^{+0.24}_{-0.26}(\mathrm{syst})$~mb.
The result is compared to theoretical models for coherent \jpsi~production and found to be in
good agreement with those models which include nuclear gluon shadowing.
\end{abstract}
\end{titlepage}
\setcounter{page}{2}
%

Two-photon and photonuclear interactions at unprecedentedly high energies can be studied in
ultra-peripheral heavy-ion collisions (UPC) at the LHC. In such collisions the nuclei are
separated by impact parameters larger than the sum of their radii and therefore hadronic
interactions are strongly suppressed. The cross sections for photon induced reactions
remain large because the strong electromagnetic field of the nucleus enhances the intensity
of the virtual photon flux, which grows as $Z^2$, where $Z$ is the charge of the nucleus.
The virtuality of the photons is restricted by the nuclear form factor to be of the order
$1/R \approx$~30~MeV/$c$ ($R$ is the radius of the nucleus).
The physics of ultra-peripheral collisions is reviewed in ~\cite{Review2008,Review2005}. 

Exclusive photoproduction of vector mesons, where a vector meson but no other particles are
produced in the event, is of particular interest.
Exclusive production of \jpsi~in photon-proton interactions,
$\gamma+ \mathrm{p} \rightarrow \jpsi + \mathrm{p}$, has been
successfully modelled in perturbative QCD in terms of the exchange of two gluons with no net-colour
transfer\ \cite{Frankfurt:1997fj}.
Experimental data on this process from HERA have been used to constrain the proton gluon-distribution at low
Bjorken-$x$\ \cite{Martin:2007sb}. Exclusive vector meson production in heavy-ion
interactions is expected to probe the nuclear gluon-distribution\ \cite{Rebyakova:2011vf}, for which there
is considerable uncertainty in the low-$x$ region\ \cite{Eskola:2009uj}.
A \jpsi~produced at rapidity $y$ is sensitive to the gluon distribution at
$x = (M_{\jpsi}/\sqrt{s_\mathrm{NN}}) \exp(\pm y)$ at hard scales
$Q^2 \approx M_{\jpsi}^2$/4\ \cite{Ryskin:1992ui}. The two-fold ambiguity in $x$ is due to the fact that 
either nucleus can serve as photon emitter or photon target. 
At the forward rapidities studied here (-3.6 $<$ $y$ $<$ -2.6), 
the relevant values of $x$ are $\approx 10^{-2}$ and $\approx 10^{-5}$, respectively.

Exclusive $\rho^0$~\cite{Abelev:2007nb} and \jpsi~\cite{Afanasiev:2009hy} production have been studied in Au-Au
collisions at RHIC. The $\rho^0$ is too light to provide a hard scale, and the \jpsi~analysis suffered from 
very low statistics, so no conclusions concerning nuclear shadowing were made from these studies. 
Exclusive \jpsi~production has also been studied by the CDF collaboration in proton-antiproton collisions at the
Tevatron\ \cite{Aaltonen:2009kg}. The availability of such measurements has led to an increase in interest in 
ultra-peripheral collisions, stimulating several new model calculations.

In this Letter, the first LHC results on exclusive photoproduction of \jpsi~vector mesons are presented.
\jpsi~mesons produced in Pb--Pb collisions at \snn~= 2.76 TeV have been measured at forward rapidities 
through their dimuon decay.
Exclusive photoproduction can be either coherent, where the
photon couples coherently to all nucleons, or incoherent, where the photon couples to a single
nucleon. Coherent production is characterized by low vector meson transverse momentum
($\langle p_{\rm T} \rangle \simeq$~60~MeV/$c$) and the target nucleus normally does not break up. Incoherent production,
corresponding to quasi-elastic scattering off a single nucleon, is characterized by a somewhat higher
transverse momentum ($\langle p_{\rm T} \rangle \simeq$~500~MeV/$c$) and the target nucleus normally breaks up, but
except for single nucleons or nuclear fragments in the very forward region no other particles are
produced. 
This analysis is focussed on coherently produced {\jpsi}~mesons. The experimental definition of coherent 
production, which must take into consideration also the finite detector resolution, is here \pt~$<$~0.3~GeV/$c$. 
The measured cross section is compared to model 
predictions~\cite{Rebyakova:2011vf,starlight,Adeluyi:2012ph,Goncalves:2011vf,Cisek:2012yt}.

The ALICE detector consists of a central barrel placed inside a large solenoid magnet ($B = 0.5$~T),
covering the pseudorapidity region $\vert\eta\vert$ $<$ 0.9~\cite{Aamodt:2008zz}, and a muon
spectrometer covering the range --4.0$<\eta<$--2.5. The spectrometer consists
of a ten interaction length ($\lambda_I$) thick absorber filtering the muons, in front of five
tracking stations containing two planes of cathode pad multi-wire proportional chambers (MWPC) each,
with the third station placed inside a dipole magnet with a $\int B dl =$~3~Tm integrated field. The forward muon
spectrometer includes a triggering system, used to select muon candidates with a transverse momentum
larger than a given programmable threshold. It has four planes of resistive plate chambers (RPC)
downstream of a 1.2 m thick iron wall (7.2 $\lambda_I$), which absorbs secondary punch-through hadrons
from the front absorber and low momentum muons from $\pi$ and K weak decays.
This analysis uses the VZERO counters for triggering and event selection. These consist of
two arrays of 32 scintillator tiles each, covering the range 2.8$<$ $\eta$ $<$5.1
(VZERO-A, on the opposite side of the muon arm) and --3.7$<$$\eta$ $<$--1.7
(VZERO-C) and positioned at $z$~= 329~cm and $z$~= --87~cm from the interaction point, respectively.
Finally, two sets of hadronic Zero-Degree Calorimeters (ZDCs) are located at 116 m on either side of
the Interaction Point. These detect neutrons emitted in the very forward region, for example neutrons 
emitted following electromagnetic dissociation\ \cite{ALICE_EMD}.  

The analysis presented in this publication is based on a sample of events collected during the
2011 \pb~run, selected with a special trigger (FUPC) set up to select UPC events in which a dimuon pair 
is produced within the acceptance of the detector. 
The integrated luminosity corresponds to about 55 $\mu b^{-1}$. 

The purpose of the FUPC trigger is to select events containing two muons 
from two-photon production ($\gamma \gamma \rightarrow \mu^{+}\mu^{-}$) or from \jpsi~decay, and it 
requires the following event characteristics: \newline
(i) a single muon trigger above a 1 GeV/$c$ \pt-threshold; \\
(ii) at least one hit in the VZERO-C detector since the muon spectrometer covers most of 
its pseudorapidity acceptance. In addition, VZERO-C vetoes the remaining upstream beam-gas events 
which could produce a trigger in the muon arm; \\
(iii) no hits in the VZERO-A detector to reject hadronic collisions. \\
A total of 3.16 $\times 10^{6}$ events were selected by the FUPC trigger.

The offline event selection used in a previous \jpsi~analysis~\cite{Aamodt:2011gj} was modified to account
for the typical experimental signatures of ultra-peripheral processes, \textit{i.e.} only two tracks in the 
spectrometer and very low \jpsi~transverse momentum.
The following selection criteria were applied (number of remaining events after the selection): \newline
(i) two reconstructed tracks in the muon arm (432,422 events); \\
(ii) owing to the multiple scattering in the front absorber, the DCA (distance between the
vertex and the track extrapolated to the vertex transverse plane) distribution of the tracks coming
from the interaction vertex can be described by a Gaussian function, whose width depends on the absorber
material and is proportional to $1/p$, where $p$ is the muon momentum. The beam induced background does not 
follow this trend, and was
rejected by applying a cut on the product $p \times \mathrm{DCA}$, at 6 times the standard deviation of the dispersion
due to multiple scattering and detector resolution. The additional dispersion due to the uncertainty on the vertex 
position (not measurable in UPC events) is negligible in comparison and does not affect the value of the cut 
(26,958 events); \\
(iii) at least one of the muon track candidates were required to match a trigger track above the 1~GeV/$c$ 
\pt-threshold in the spectrometer trigger chambers (10,172 events); \\
(iv) both tracks pseudorapidities within the range
--3.7$<\eta_{1,2}<$--2.5, to match the VZERO-C acceptance (5,100 events); \\
(v) the tracks exit from the absorber in the range 17.5~cm$<R_{\mathrm{abs}}<$89.5~cm, delimiting the two
homogeneous parts of the absorber covering the angular acceptance of the spectrometer ($R_\mathrm{abs}$
is the radial coordinate of the track at the end of the front absorber) (5,095 events); \\
(vi) dimuon rapidity to be in the range --3.6$< y <$--2.6,
which ensured that the edges of the spectrometer acceptance were avoided (4,919 events); \\
(vii) two tracks with opposite charges (3,209 events); \\
(viii) only events with a neutron ZDC signal below 6 TeV on each side were kept. In the present
data sample, this cut does not remove any events with a \jpsi~produced with a transverse momentum
below 0.3 GeV/$c$, but reduces hadronic contamination at higher \pt~(817 events); \\
(ix) dimuons to have \pt~$<$~0.3~GeV/$c$ and invariant mass 2.8 $<$ \minv~$<$ 3.4~GeV/$c^2$ (122 events); \\
(x) VZERO offline timing compatible with crossing beams (117 events). \\

The acceptance and efficiency of \jpsi-reconstruction were calculated using a large sample of
coherent and incoherent \jpsi~events generated by STARLIGHT\ \cite{starlightMC} and
folded with the detector Monte Carlo simulation. STARLIGHT simulates 
photonuclear and two-photon interactions at hadron colliders. The simulations
for exclusive vector meson production and two-photon interactions are based on the models
in \cite{starlight} and \cite{Baltz:2009jk}, respectively. 

The residual misalignment and the time-dependent conditions of the
tracking and trigger chamber components were taken into account in these simulations.
The trigger chamber efficiencies were computed from the data and used in the global efficiency
calculation. A separate simulation was performed for each run, in order to take into account the slight
variations in run conditions during the data taking.
The product of the acceptance and efficiency corrections $\acceps$~was calculated as
the ratio of the number of the simulated events that satisfy the event selection in Table~1 to the
number of generated events within --3.6~$< y <$~--2.6.
The final values for the combined acceptance and efficiency were found to be 16.6\% and 14.3\% for 
coherent and incoherent \jpsi, respectively.
The relative systematic error coming from the uncertainties on
the trigger chamber efficiencies used in these simulations amounts to 4\%. In addition, the muon
reconstruction efficiency has been evaluated both in data and simulations, in a way similar to that
described in\ \cite{Aamodt:2011gj}, and a 6\% relative systematic uncertainty on the $\acceps$~corrections
was assigned to account for the observed differences.

In order to evaluate the systematic error on the acceptance coming from the generator choice,
the acceptance was computed from a parameterization of the results on coherent \jpsi~production
in \cite{Rebyakova:2011vf}. It was also calculated by modifying the rapidity distribution in STARLIGHT
and letting it vary between a flat distribution and a distribution consistent with the model with the
steepest slope (AB-MSTW08, see below for definition) over the range --3.6~$< y <$~--2.6.
The differences in acceptance between the methods were below 3\%, which was
taken into account in the systematic error calculation.
It is assumed in these calculations that the \jpsi~is transversely
polarized. Transverse polarization
is expected for a quasi-real photon from s-channel helicity conservation. This has been confirmed
experimentally for exclusive \jpsi~production in $\gamma+ {\rm p} \rightarrow \jpsi + {\rm p}$
interactions \cite{Chekanov:2002xi,Aktas:2005xu} and for exclusive $\rho^0$ photoproduction in
heavy-ion collisions \cite{Abelev:2007nb}. Owing to the low \pt~of the \jpsi , the calculations are
insensitive to the choice of reference frame (here the helicity frame was used), and the polarization
axis effectively coincides with the beam axis.

Activity in the central barrel was checked for events with invariant mass in the range
2.8  $<$ \minv~$<$ 3.4~GeV/$c^2$. No events with more than one tracklet in the Si-Pixel (SPD) detector
were found. The events with one tracklet (6 out of 117) were not removed, as this level of activity
is consistent with the background from random combinations of noise hits. 

The invariant mass distribution for  opposite sign (OS) muon pairs with
2.2 $<$ \minv~$<$ 4.6 GeV/$c^{2}$ is shown in Fig. 1. A \jpsi~peak is clearly
visible in the spectrum, on top of a continuum coming from
$\gamma \gamma \rightarrow\mu^{+}\mu^{-}$. Only two like--sign dimuon pairs are in the invariant mass range
2.2 $<$ \minv~$<$ 4.6 GeV/$c^{2}$, at 2.3 GeV/$c^{2}$ and 2.8 GeV/$c^{2}$. 
The combinatorial background is therefore estimated to be $\leq$2\% at 90\% confidence level in the 
invariant mass range 2.8 $<$ \minv~$<$ 3.4 GeV/$c^{2}$.

The \jpsi~yield was obtained by fitting the dimuon invariant mass spectrum in the
range 2.2 $<$ \minv~$<$ 4.6 GeV/$c^{2}$ with an exponential function
to describe the underlying continuum, and a Crystal Ball function\ \cite{Gai:CrystalBall}
to extract the \jpsi~signal. The Crystal Ball tail parameters ($\alpha_{CB}$ and $n$) were fixed to values
obtained from simulations. 
The central mass value from the fit is $3.123 \pm 0.011$~GeV/$c^2$, which is within 2.4$\sigma$ (0.8\%)
of the known value of the \jpsi~mass and compatible with the absolute calibration accuracy of the muon
spectrometer.
The width, 84$\pm$14 MeV/$c^{2}$, is in agreement with the Monte Carlo simulations. The extracted number of
{\jpsi}s~is $N_{\rm yield} = 96 \pm 12$(stat) $\pm$ 6(syst). The systematic error on the yield (6\%) 
was obtained by varying the Crystal Ball tail parameters. The exponential slope parameter of the
continuum is -1.4$\pm$0.2~GeV$^{-1} c^2$ in good agreement with the corresponding Monte Carlo
expectation (-1.39$\pm$0.01~GeV$^{-1} c^2$). This, together with the fact that the \pt~distribution is 
consistent with the expectations from STARLIGHT, is an additional indication that there is no unexpected 
background in the invariant mass range considered. 

The fraction $f_D$ of the \jpsi~mesons coming from the decay of $\psi^{'} \rightarrow \jpsi$~+~anything was estimated by
simulating a sample of coherently produced {$\psi^{'}$}s with STARLIGHT, using PYTHIA\ \cite{Pythia} to simulate 
their decay into \jpsi. The detector response was simulated as described above. The contribution from incoherently 
produced $\psi^{'}$ is expected to give a negligible contribution for \pt~$<$~0.3~GeV/$c$ and was not considered.
Unlike the directly produced \jpsi~discussed above, the polarization of {\jpsi}s~coming from $\psi^{'}$ decays 
cannot easily be predicted, since the polarization of the original $\psi^{'}$ can be shared between the \jpsi~and 
the other daughters in different ways. The $\psi^{'}$ decay was therefore simulated by assuming
the following \jpsi~polarizations: (i) no polarization (NP); (ii) full transverse (T), and
(iii) full longitudinal (L).
The \jpsi~fraction coming from  $\psi^{'}$ decay for a given
polarization $P$, $f_{D}^{P}$\ can be written as:

\begin{equation}
f_{D}^{P}
=\frac
{\sigma_{\psi^{'}}  \cdot
BR(\psi^{'} \rightarrow \jpsi+\mathrm{anything}) \cdot
{(\mathrm{ Acc}\times\varepsilon)}_{\psi^{'}\rightarrow\jpsi}^{P}  }
{\sigma_{\jpsi} \cdot
\acceps} \; ,
\label{eq1a2}
\end{equation}
where the \acceps~and $(\mathrm{ Acc}\times\varepsilon)_{\psi^{'}\rightarrow\jpsi}^{P}$ were
computed for \pt~$<$ 0.3 GeV/$c$.

According to STARLIGHT, the ratio between the $\psi^{'}$ and $J/\psi$ coherent photoproduction
cross sections is 0.19 giving
$f_{D}^{NP} =$ 11.9$\%$,
$f_{D}^{T} =$ 9.3$\%$,
$f_{D}^{L} =$ 16.8$\%$.
The cross sections ratio is significantly lower in the pQCD inspired model
\cite{Rebyakova:2011vf}, 0.087.
This changes the above fraction, giving
$f_{D}^{NP} =$ 5.5$\%$,
$f_{D}^{T} =$ 4.3$\%$,
$f_{D}^{L} =$ 7.9$\%$.
The estimates for $f_D$ thus range from 4.3$\%$ to 16.8$\%$. The best estimate was taken 
as the middle of this range with the extremes providing the lower and upper limits, 
giving $f_{D}$=(11$\pm$6)$\%$.

The dimuon \pt~distribution integrated over 2.8 $<$ \minv~$<$ 3.4~GeV/$c^2$ is presented in
Fig. 2. 
The clear peak at
low \pt~is mainly due to coherent interactions, while the tail extending out to 0.8 GeV/$c$
comes from incoherent production. In addition, the high-\pt~region may still contain a few
hadronic events, which makes it difficult to extract the incoherent photoproduction cross section
from these data.
To estimate the fraction ($f_I$) of incoherent over coherent events in the
region \pt~$<$~0.3~GeV/$c$, the ratio $\sigma_\mathrm{inc}/\sigma_\mathrm{coh}$, weighted by the detector
acceptance and efficiency for the two processes, was calculated, giving $f_{I} = 0.12$ when
$\sigma_\mathrm{inc}/\sigma_\mathrm{coh}$ was taken from STARLIGHT, and $f_{I} = 0.08$ when the
model in \cite{Rebyakova:2011vf} was used.
Four different functions were used to describe the \pt~spectrum: coherent and incoherent
photoproduction of \jpsi, \jpsi~from $\psi^{'}$ decay, and two-photon production of continuum
pairs. The shapes for the fitting functions (Monte Carlo templates) were provided by
STARLIGHT events folded with the detector simulation. The relative normalization was left
free for coherent and incoherent photoproduction.
The contribution from the $\psi^{'}$ was constrained from estimate above ($f_{D}$=(11$\pm$6)$\%$), 
and the two-photon contribution was determined from the fit of the
continuum in Fig.~1. 
In the fit, the incoherent process is constrained  mainly in the region 0.5~$<$~\pt~$<$~0.8 GeV/$c$,
where the other three processes are negligible. As this \pt~region (not used in the \jpsi~signal
extraction) is likely to suffer from some hadronic background,
the fit can only provide an upper limit on $f_I$. The result is $f_I = 0.26 \pm 0.05$, about
a factor 2 larger than the estimate from the theoretical models quoted above.
We conclude by taking the middle value of the two calculations and the fit as the best 
estimate of $f_I$, and the other two results as lower and upper limits, respectively, 
giving $f_{I} = 0.12^{+0.14}_{-0.04}$.

The fact that the Monte Carlo templates describe the \pt~distribution well in the range 0.0~$<$~\pt~$<$~0.8 GeV/$c$ 
confirms that there is no strong contamination from hadronic production in the event sample. An upper 
limit on the contribution from hadronic interactions can be obtained 
by considering events with \pt~$>$~1.0~GeV/$c$, where the contribution from incoherent photoproduction is very small. 
For hadronic \jpsi~production it is known from the parameterization in Ref.~\cite{Bossu:2011qe} that (including the 
acceptance and efficiency corrections) 82\% of 
the yield is above \pt~$>$~1.0~GeV/$c$, while only 2\% is below \pt~$<$~0.3~GeV/$c$. If one conservatively 
assumes that the 32 events in 
the data sample with \pt~$>$~1.0~GeV/$c$ are all from hadronic production, the expected yield from hadronic 
interactions below \pt~$<$~0.3~GeV/$c$ can be estimated to be $(0.02/0.83) \cdot 32 = 0.8$ events. This is 
thus less than a 1\% contamination. A similar estimate can be 
obtained by scaling the measured cross section for \jpsi~production in Pb-Pb collisions~\cite{Aamodt:2011gj} with the 
number of binary collisions assuming that all events with 80-100\% centrality survive the event selection (a very 
conservative assumption). The conclusion is thus that the contamination from hadronic interactions is negligible 
and no correction need be applied for it.

Finally, the total number of coherent \jpsi s is calculated from the yield extracted from the fit to the
invariant mass distribution by
\begin{equation}
N^{\mathrm{coh}}_{\jpsi} = \frac{N_{\rm yield}}{1 + f_I + f_D} \; ,
\label{NCohJPsi}
\end{equation}
resulting in $N^{\mathrm{coh}}_{\jpsi}  = 78 \pm 10(\mathrm{stat}) ^{+7}_{-11} (\mathrm{syst}) $.

The coherent \jpsi~differential cross section is given by:

\begin{equation}
\frac{\mathrm{d}\sigma_{\jpsi}^{\mathrm{coh}}}{\mathrm{d}y}  =
\frac{N_{\jpsi}^{\mathrm{coh}}}
{\acceps \cdot \epsilon_{\mathrm{trig}} \cdot BR(\jpsi \longrightarrow \mu^{+}\mu^{-}) \cdot {\cal{L}_{\mathrm{int}}} \cdot\Delta y  } \; ,
\label{eq1a}
\end{equation}

where $N^{\mathrm{coh}}_{\jpsi}$ is the number of \jpsi~candidates from Eq.~\ref{NCohJPsi},
$\acceps$~corresponds to the acceptance and efficiency of the muon spectrometer, as discussed above,
and $\epsilon_{\mathrm{trig}}$ is the VZERO trigger efficiency.
$BR(\jpsi \longrightarrow \mu^{+}\mu^{-}) =$~5.93\% is the branching ratio for \jpsi~decay into muons\ \cite{Beringer:1900zz},
$\Delta y =$~1 the rapidity interval bin size, and $\cal{L}_{\mathrm{int}}$
the total integrated luminosity.
During the 2011 Pb-Pb run the VZERO detector was optimised for the
selection of hadronic Pb-Pb collisions, with a threshold corresponding to an
energy deposit above that from a single minimum ionizing particle (MIP).
The distribution of the signal produced by a MIP crossing the 2 cm thick VZERO
scintillator has a Landau shape. To get an accurate simulation of the efficiency for 
low multiplicity events with this threshold setting, would require an almost perfect
reproduction of the Landau by the MC simulation. Therefore we used the QED continuum
pair production for the normalization and not Eq.~3.

In addition to exclusive \jpsi , the FUPC trigger selected $\gamma \gamma \rightarrow \mu^+ \mu^-$ events,
which are very similar to coherent \jpsi~decays in terms of kinematics and associated event characteristics.
This reaction is a standard QED process, which in principle can be calculated with high accuracy. The fact
that the photon coupling to the nuclei is $Z \sqrt{\alpha}$ (with Z = 82 here) rather than just 
$\sqrt{\alpha}$ increases the uncertainty
of the contribution from higher order terms. Predictions exist where this effect
is negligible\ \cite{Hencken:2006ir}. However, other studies obtained a 16\% reduction in the cross section
from higher order terms in Pb-Pb collisions at the LHC\ \cite{Baltz:2009fs}. There is also an uncertainty
associated with the minimum momentum transfer and the nuclear form factor\ \cite{Baltz:2010mc}.
Two-photon production of $\mu^+ \mu^-$--pairs from
STARLIGHT was used to determine the trigger efficiency\ \cite{Baltz:2009jk}. The cross sections from STARLIGHT
for two-photon production
of $\mathrm{e}^+ \mathrm{e}^-$ and $\mu^+ \mu^-$\ pairs have previously been compared with results from
STAR\ \cite{Adams:2004rz} and PHENIX\ \cite{Afanasiev:2009hy}, respectively. The predictions from
STARLIGHT have been found to be in good agreement with the experimental results. These results, however,
have uncertainties of about 25 to 30\%. In the absence of high precision measurements constraining the
model, and taking into account the outstanding theoretical issues mentioned above, the uncertainty in the
STARLIGHT two-photon cross section is estimated to be 20\%.

The cross section for $\gamma \gamma \rightarrow \mu^+ \mu^-$ can be written in a similar
way to Eq.~3 and the ratio of the two is independent of luminosity and of the trigger efficiency:

\begin{equation}
\frac{\mathrm{d}\sigma_{\jpsi}^{\mathrm{coh}}}{\mathrm{d} y}   =
\frac{1}{BR(\jpsi \rightarrow \mu^+ \mu^-)}
\cdot
\frac{N^{\mathrm{coh}}_{\jpsi}}{N_{\gamma\gamma}}
\cdot
\frac{(\mathrm {Acc}\times\varepsilon)_{\gamma\gamma} }
{\acceps}
\cdot
\frac{\sigma_{\gamma\gamma}}{\Delta y} ,
\label{comp2}
\end{equation}

where $N_{\gamma\gamma}$ was obtained by counting the number of events in the invariant mass intervals 
2.2  $<$ \minv~$<$ 2.6~GeV/$c^{2}$ ($N_{\gamma\gamma} =$~43$\pm$7(stat.)) and
3.5 $<$ \minv~$<$ 6~GeV/$c^{2}$ ($N_{\gamma\gamma} =$~15$\pm$4(stat.)), to avoid contamination from the \jpsi~peak.
To determine  $\sigma_{\gamma \gamma}$ STARLIGHT\ \cite{Baltz:2009jk} was used. The cross section
for dimuon invariant mass between 2.2  $<$ \minv~$<$ 2.6~GeV/$c^{2}$ or 3.5 $<$ \minv~$<$ 6~GeV/$c^{2}$, 
dimuon rapidity\ in the interval -3.6 $< y <$ -2.6, and each muon satisfying
-3.7 $< \eta_{1,2} <$ -2.5 is $\sigma_{\gamma \gamma} = 17.4$~$\mu$b ($\sigma_{\gamma \gamma} = 13.7$~$\mu$b and
$\sigma_{\gamma \gamma} = 3.7$~$\mu$b for the low and high invariant mass intervals, respectively). 
The $(\mathrm{Acc}\times\varepsilon)_{\gamma\gamma}$ for events satisfying the same selection was calculated 
using events from STARLIGHT folded with the detector simulation as described above.
The data cuts applied to the Monte Carlo sample were the same as those applied for
the \jpsi~data analysis, resulting in a $(\mathrm{Acc}\times\varepsilon)_{\gamma\gamma}$ of 42.1\% 
(37.9\% for 2.2 $<$ \minv~$<$ 2.6~GeV/$c^{2}$ and 57.5\% for 3.5 $<$ \minv~$<$ 6~GeV/$c^{2}$). 

\begin{table}
\caption{\label{tab:syserr} Summary of the contributions to the systematic uncertainty for
the integrated \jpsi~cross section measurement. The error for the coherent signal extraction includes
the systematic error in the fit of the invariant mass spectrum and the systematic errors on
$f_D$ and  $f_I$, as described in the text.}	
\begin{center}
\begin{tabular}{lcc}
Source                                                & Value   	 \\ \hline	
Theoretical uncertainty in $\sigma_{\gamma \gamma}$    & 20\%              \\
Coherent signal extraction                            &  $^{+9}_{-14}$\%  \\
Reconstruction efficiency             		      &  6\% 	         \\
RPC trigger efficiency           		      &  5\% 		 \\ 		
\jpsi~acceptance calculation    		      &  3\%             \\
Two-photon e$^+$e$^-$ background 		      &  2\%             \\
Branching ratio 				      &  1\%             \\ \hline
Total                                                 &  $^{+24}_{-26}$\%  \\
\end{tabular}
\end{center}
\end{table}

A possible source of inefficiency comes from correlated QED pair production, \textit{i.e.}
interactions which produce both a \jpsi~and a low mass $e^+ e^-$--pair (the latter has a
very large cross section), with one of the electrons hitting the VZERO-A detector and thus
vetoing the event.
This effect was studied with data, in a sample collected with comparable luminosity
by a control trigger, requiring a coincidence
of at least two muons in the muon arm trigger with hits in both the VZERO-A and VZERO-C.
Two \jpsi~events were found in this sample, giving an upper limit on the inefficiency smaller than 2\%.

Since the kinematic distributions of the muons  from \jpsi~decays and $\gamma\gamma$ 
processes are different, the systematic uncertainties on the corresponding $(\mathrm{Acc}\times\varepsilon)$
corrections coming from the uncertainties on the muon trigger and reconstruction efficiencies do not exactly
cancel out in equation\ \ref{comp2}. In order to account for this effect, a 50\% correlation factor has
been estimated, conservatively, when computing the systematic uncertainty on the ratio.
The sources of the systematic error are summarized in  Table~\ref{tab:syserr}. 
The final result is a differential cross section for coherent \jpsi~production of
$\mathrm{d}\sigma_{J/\psi}^{\mathrm{coh}} /\mathrm{d}y = 1.00 \pm 0.18(\mathrm{stat}) ^{+0.24}_{-0.26}(\mathrm{syst})$~mb.

The cross section is compared with
 calculations from various models\ \cite{Rebyakova:2011vf,starlight,Adeluyi:2012ph,Goncalves:2011vf,Cisek:2012yt}
in Fig. 3. 
The differences between the
models come mainly from the way the photonuclear interaction is treated. The predictions can be divided into
three categories: \newline
i) those that include no nuclear effects (AB-MSTW08, see below for definition). In this
approach, all nucleons contribute to the scattering, and the forward scattering differential cross section,
$\mathrm{d} \sigma/\mathrm{d} t$ at $t = 0$ ($t$ is the momentum transfer from the target nucleus squared),
scales with the number of nucleons squared, $A^2$; \\
ii) models that use a Glauber approach to calculate the number of nucleons
contributing to the scattering (STARLIGHT, GM, and CSS). The reduction in the calculated cross section
depends on the total \jpsi-nucleon cross section; \\
iii) partonic models, where the cross section is proportional to the nuclear gluon
distribution squared (AB-EPS08, AB-EPS09, AB-HKN07, and RSZ-LTA).

STARLIGHT uses the latest HERA data on exclusive \jpsi~production in photon-proton interactions\cite{Chekanov:2002xi,Aktas:2005xu} 
as input to calculate the corresponding photon-nucleus cross section. 
The model by Goncalves and Machado (GM)\ \cite{Goncalves:2011vf} calculates the \jpsi-nucleon cross section 
from the Color Dipole model, whereas Cisek, Szczurek, and Schafer (CSS)\ \cite{Cisek:2012yt} use the essentially 
equivalent $k_\perp$-factorization approach. The difference of about 25\%
between the two calculations is due to different treatment of the nucleon gluon distribution at
low $x$ (gluon saturation), and the way in which it affects the dipole-nucleon cross section.

Calculations by Adeluyi and Bertulani (AB)\ \cite{Adeluyi:2012ph} and by Rebyakova, Strikman, and
Zhalov (RSZ)\ \cite{Rebyakova:2011vf} are based on perturbative QCD. 
The calculations by Rebyakova {\it et al.} use a cross section for exclusive \jpsi~photoproduction on a
proton target calculated from leading order perturbative QCD within the leading log approximation. The
calculations use the integrated gluon density distribution in the proton determined by the
Durham-PNPI group from data on exclusive \jpsi~production at HERA\ \cite{Martin:2007sb}. 
The modification to the nuclear gluon distribution has been calculated in the Leading Twist 
Approximation\ \cite{Frankfurt:1998ym} and is based on using the DGLAP evolution equations and the 
HERA diffractive parton density distributions. 

Adeluyi and Bertulani constrain the nucleon parton distributions to be consistent with data on exclusive
vector meson production in photon-proton interactions. The photonuclear cross section is then calculated
using different standard parameterizations of the nuclear gluon distribution functions
(EPS08, EPS09, and HKN07). For comparison, they also performed calculations 
where the constrained nucleon gluon distribution function is scaled with the number of nucleons without 
shadowing or other nuclear effects (AB-MSTW08).

In the region of interest here, --3.6$< y <$--2.6, the sensitivity to shadowing is reduced
compared with that at mid-rapidity.
Away from mid-rapidity, there is a two-fold
ambiguity in the photon energy and the momentum transfer from the nucleus acting as photon target.
For example, a \jpsi~produced at $y = 3$ corresponds to a photon-proton center-of-mass energy of
either $W_{\gamma\mathrm{p}} = 414$~GeV or $W_{\gamma\mathrm{p}} = 21$~GeV. These two energies in turn correspond 
to values of $x$ of
about $5 \times 10^{-5}$ and $2 \times 10^{-2}$, respectively. According to STARLIGHT
interactions with $W_{\gamma\mathrm{p}} = 21$~GeV contribute 94\% of the cross section, while events
with $W_{\gamma\mathrm{p}} = 414$~GeV contribute only 6$\%$. The total
$\mathrm{d}\sigma_{J/\psi}^{\mathrm{coh}} /\mathrm{d}y$ at $y$ = 3 is therefore
mainly sensitive to the gluon distribution around $x = 2 \times 10^{-2}$.

The measured cross section, $\mathrm{d}\sigma_{J/\psi}^{\mathrm{coh}} 
/\mathrm{d}y = 1.00 \pm 0.18(\mathrm{stat}) 
^{+0.24}_{-0.26}(\mathrm{syst})$~mb, is compared with the model 
predictions in Fig. 3 a). Fig. 3 b) shows a comparison of the cross 
section integrated over the range --3.6$< y <$--2.6. The models with largest 
deviations from the measured value are STARLIGHT and AB-MSTW08, 
which both deviate by about 3 standard deviations if the statistical and 
systematic errors are added in quadrature. Best agreement (within one standard 
deviation) is seen for the models
RSZ-LTA, AB-EPS09, and AB-EPS08, which include nuclear gluon shadowing. 
A further check can be performed by 
dividing the rapidity interval in two and determining the ratio of the 
cross sections in each interval. This has the advantage that some parts of the 
systematic errors cancel, and the dominant remaining error is the 
statistical error. The result is $R = \sigma(-3.1<y<-2.6)/ 
\sigma(-3.6<y<-3.1) = 1.36 \pm 0.36(\mathrm{stat}) \pm 0.19 
(\mathrm{syst})$. The systematic error includes the uncertainties in the
signal extraction and in the trigger and recontruction efficiency. 
The measured ratio is compared with that from the models in Fig. 
3 c). The only models which deviate by more than one standard deviation are 
AB-MSTW08 and AB-HKN07 (1.7 and 1.5 standard deviations, respectively).

In summary, the first LHC measurement on exclusive photoproduction of 
\jpsi~in \pb-collisions at \snn~= 2.76 TeV has been presented and 
compared with model calculations. The AB-MSTW08 model, which assumes that 
the forward scattering cross section scales with the number of nucleons 
squared, disagrees with the measurement, both for the value of the cross 
section and for the ratio of the two rapidity intervals, and is strongly 
disfavoured. STARLIGHT deviates by nearly three standard deviations in the 
cross section and is also disfavoured. Best agreement is found with models 
which include nuclear gluon shadowing consistent with the EPS09 or EPS08  
parameterizations (RSZ-LTA, AB-EPS09, and AB-EPS08).

\begin{figure}[htbp]
\centering
{\includegraphics[width=0.8\linewidth,keepaspectratio]{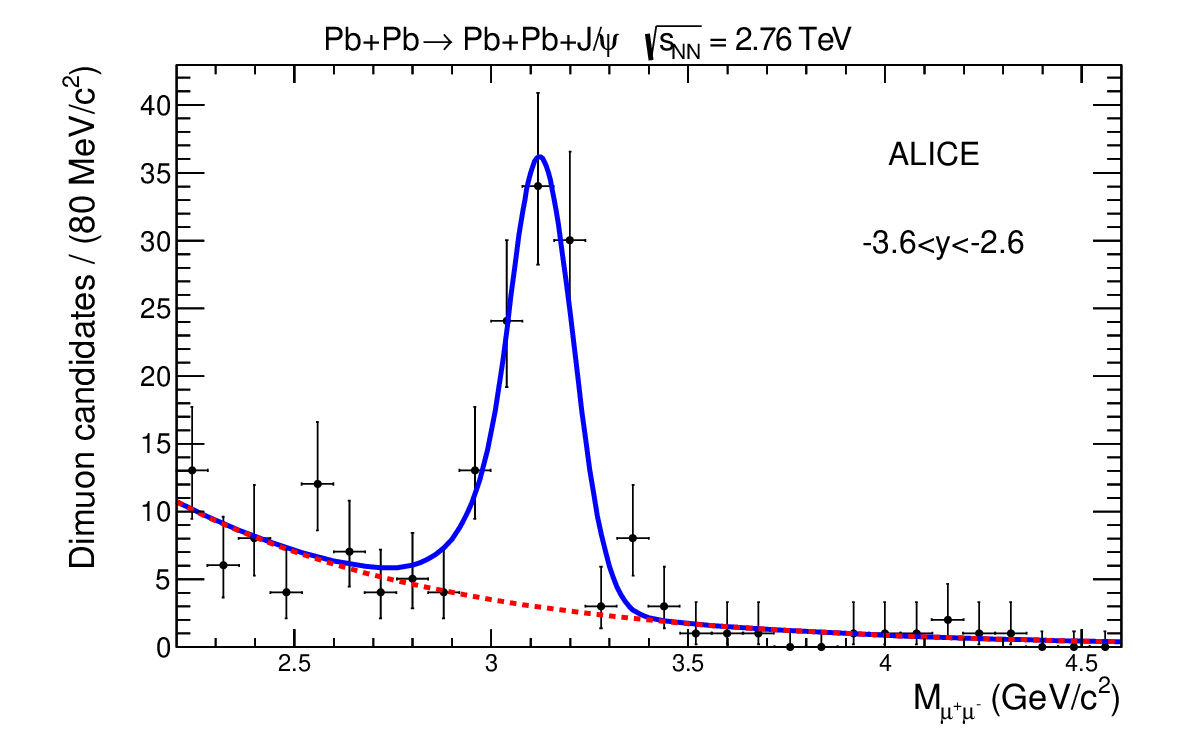}}
\caption{\label{fig:peak}Invariant mass distribution for events with exactly two oppositely charged muons 
satisfying the event selection described in the text.} 
\label{fig:1}
\end{figure}

\begin{figure}[htbp]
\begin{center}
{\includegraphics[width=0.8\linewidth,keepaspectratio]{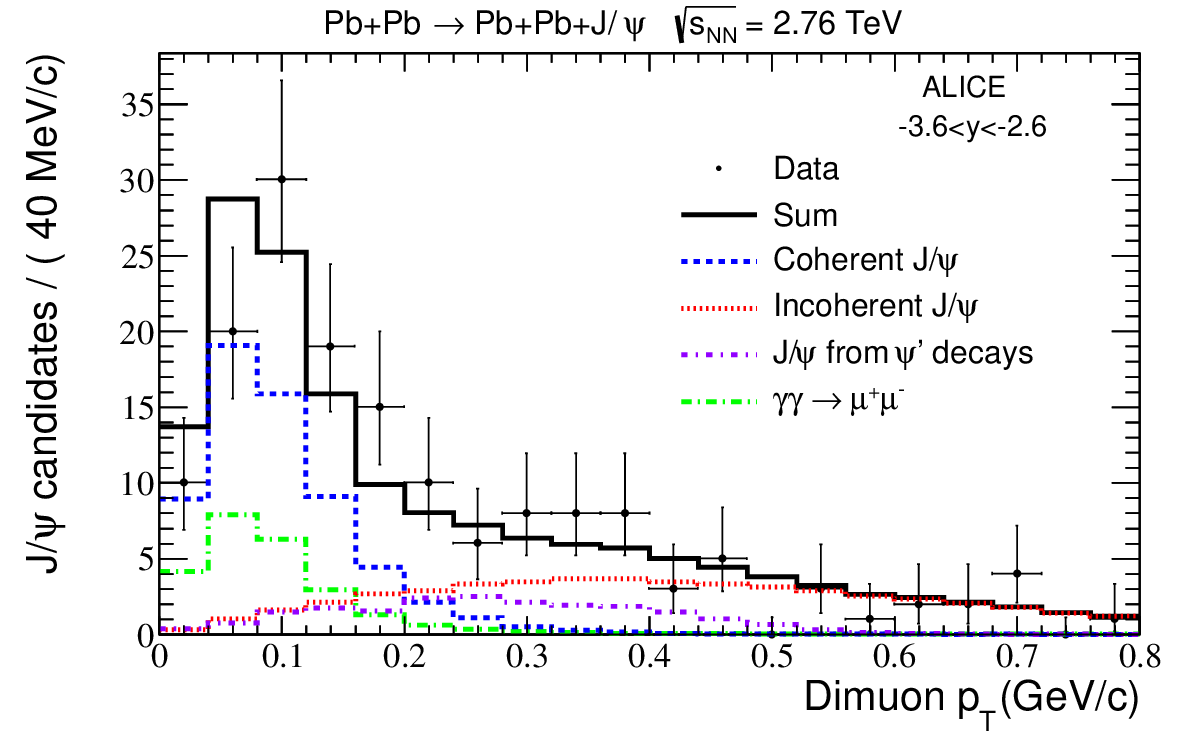}}
\end{center}
\caption{\label{fig:ptpeak} Dimuon \pt~distribution for events satisfying the event selection described in 
the text. The data points are fitted summing four different Monte Carlo templates: coherent \jpsi~production 
(dashed - blue), incoherent \jpsi~production (dotted - red), {\jpsi}s~from $\psi^{'}$ decay 
(dash-dotted - violet), and $\gamma \gamma \rightarrow \mu^+ \mu^-$ (dash-dotted - green). The solid 
histogram (black) is the sum.}
\label{fig:2}
\end{figure}

\begin{figure}
\begin{center}
\includegraphics[width=0.5\linewidth,keepaspectratio]{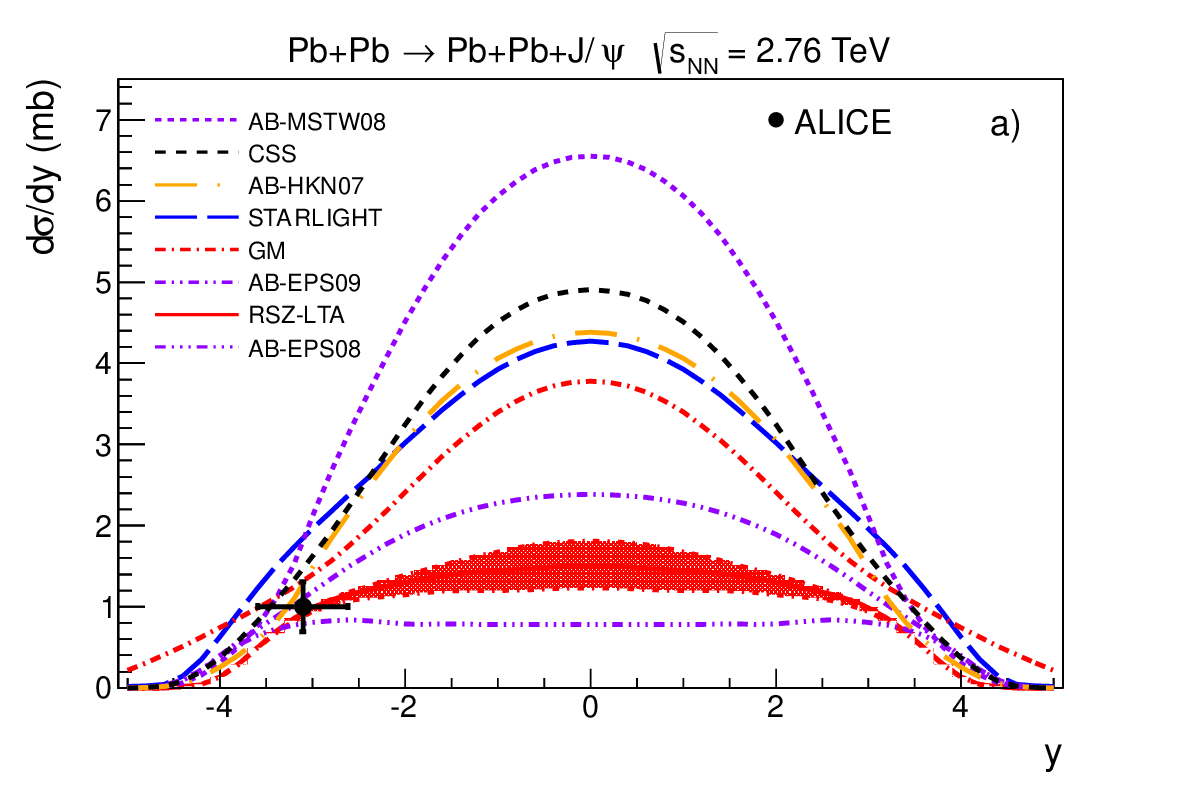}
\includegraphics[width=0.5\linewidth,keepaspectratio]{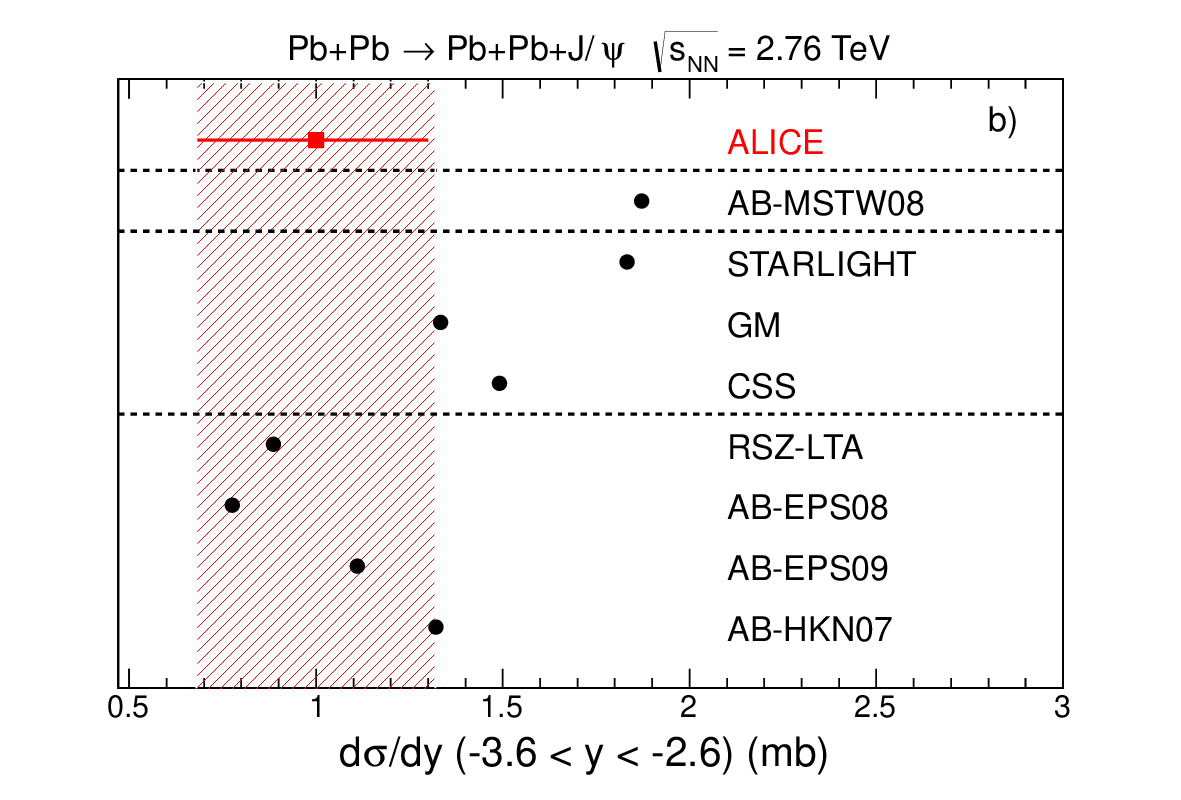}
\includegraphics[width=0.5\linewidth,keepaspectratio]{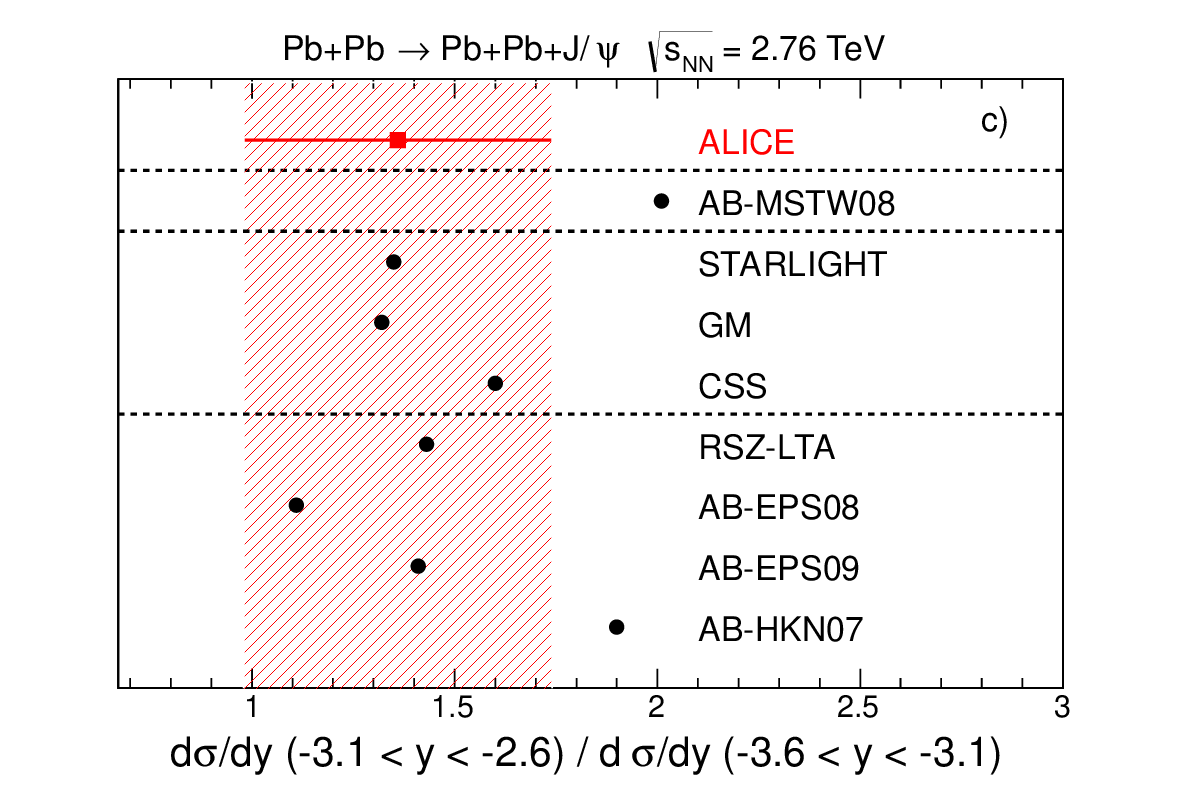}
\end{center}
\caption{\label{fig:theoryComparison} Measured coherent differential cross section of \jpsi~photoproduction in
ultra-peripheral Pb-Pb collisions at $\sqrt{s_{\mathrm{NN}} } = 2.76$ TeV. The error is the quadratic sum of the
statistical and systematic errors. The theoretical calculations described in the text are also shown. The
rapidity distributions are shown in a), b) shows the cross section integrated over -3.6 $< y <$ -2.6, and
c) shows the ratio of the cross sections in the rapidity intervals -3.1 $< y <$ -2.6 and  -3.6 $< y <$ -3.1.
The dashed lines in the lower two plots indicate the three model categories discussed in the text.}
\end{figure}




%
\newenvironment{acknowledgement}{\relax}{\relax}
\begin{acknowledgement}
\section{Acknowledgements}
The ALICE collaboration would like to thank all its engineers and technicians for their invaluable contributions to the construction of the experiment and the CERN accelerator teams for the outstanding performance of the LHC complex.
\\
The ALICE collaboration acknowledges the following funding agencies for their support in building and
running the ALICE detector:
 \\
Calouste Gulbenkian Foundation from Lisbon and Swiss Fonds Kidagan, Armenia;
 \\
Conselho Nacional de Desenvolvimento Cient\'{\i}fico e Tecnol\'{o}gico (CNPq), Financiadora de Estudos e Projetos (FINEP),
Funda\c{c}\~{a}o de Amparo \`{a} Pesquisa do Estado de S\~{a}o Paulo (FAPESP);
 \\
National Natural Science Foundation of China (NSFC), the Chinese Ministry of Education (CMOE)
and the Ministry of Science and Technology of China (MSTC);
 \\
Ministry of Education and Youth of the Czech Republic;
 \\
Danish Natural Science Research Council, the Carlsberg Foundation and the Danish National Research Foundation;
 \\
The European Research Council under the European Community's Seventh Framework Programme;
 \\
Helsinki Institute of Physics and the Academy of Finland;
 \\
French CNRS-IN2P3, the `Region Pays de Loire', `Region Alsace', `Region Auvergne' and CEA, France;
 \\
German BMBF and the Helmholtz Association;
\\
General Secretariat for Research and Technology, Ministry of
Development, Greece;
\\
Hungarian OTKA and National Office for Research and Technology (NKTH);
 \\
Department of Atomic Energy and Department of Science and Technology of the Government of India;
 \\
Istituto Nazionale di Fisica Nucleare (INFN) of Italy;
 \\
MEXT Grant-in-Aid for Specially Promoted Research, Ja\-pan;
 \\
Joint Institute for Nuclear Research, Dubna;
 \\
National Research Foundation of Korea (NRF);
 \\
CONACYT, DGAPA, M\'{e}xico, ALFA-EC and the HELEN Program (High-Energy physics Latin-American--European Network);
 \\
Stichting voor Fundamenteel Onderzoek der Materie (FOM) and the Nederlandse Organisatie voor Wetenschappelijk Onderzoek (NWO), Netherlands;
 \\
Research Council of Norway (NFR);
 \\
Polish Ministry of Science and Higher Education;
 \\
National Authority for Scientific Research - NASR (Autoritatea Na\c{t}ional\u{a} pentru Cercetare \c{S}tiin\c{t}ific\u{a} - ANCS);
 \\
Federal Agency of Science of the Ministry of Education and Science of Russian Federation, International Science and
Technology Center, Russian Academy of Sciences, Russian Federal Agency of Atomic Energy, Russian Federal Agency for Science and Innovations and CERN-INTAS;
 \\
Ministry of Education of Slovakia;
 \\
Department of Science and Technology, South Africa;
 \\
CIEMAT, EELA, Ministerio de Educaci\'{o}n y Ciencia of Spain, Xunta de Galicia (Conseller\'{\i}a de Educaci\'{o}n),
CEA\-DEN, Cubaenerg\'{\i}a, Cuba, and IAEA (International Atomic Energy Agency);
 \\
Swedish Research Council (VR) and Knut $\&$ Alice Wallenberg
Foundation (KAW);
 \\
Ukraine Ministry of Education and Science;
 \\
United Kingdom Science and Technology Facilities Council (STFC);
 \\
The United States Department of Energy, the United States National
Science Foundation, the State of Texas, and the State of Ohio.
\end{acknowledgement}
\newpage
%
%
\appendix
\section{The ALICE Collaboration}
\label{app:collab}

\begingroup
\small
\begin{flushleft}
B.~Abelev\Irefn{org1234}\And
J.~Adam\Irefn{org1274}\And
D.~Adamov\'{a}\Irefn{org1283}\And
A.M.~Adare\Irefn{org1260}\And
M.M.~Aggarwal\Irefn{org1157}\And
G.~Aglieri~Rinella\Irefn{org1192}\And
M.~Agnello\Irefn{org1313}\And
A.G.~Agocs\Irefn{org1143}\And
A.~Agostinelli\Irefn{org1132}\And
S.~Aguilar~Salazar\Irefn{org1247}
% ...
\renewcommand\labelenumi{\textsuperscript{\theenumi}~}
\section*{Affiliation notes}
\renewcommand\theenumi{\roman{enumi}}
\begin{Authlist}
\item \Adef{0}Deceased
\item \Adef{M.V.Lomonosov Moscow State University, D.V.Skobeltsyn Institute of Nuclear Physics, Moscow, Russia}Also at: M.V.Lomonosov Moscow State University, D.V.Skobeltsyn Institute of Nuclear Physics, Moscow, Russia
\item \Adef{University of Belgrade, Faculty of Physics and "Vinca" Institute of Nuclear Sciences, Belgrade, Serbia}Also at: University of Belgrade, Faculty of Physics and "Vin\v{c}a" Institute of Nuclear Sciences, Belgrade, Serbia
\end{Authlist}
\section*{Collaboration Institutes}
\renewcommand\theenumi{\arabic{enumi}~}
\begin{Authlist}
\item \Idef{org1279}Benem\'{e}rita Universidad Aut\'{o}noma de Puebla, Puebla, Mexico
\item \Idef{org1220}Bogolyubov Institute for Theoretical Physics, Kiev, Ukraine
\item \Idef{org1262}Budker Institute for Nuclear Physics, Novosibirsk, Russia
\item \Idef{org1292}California Polytechnic State University, San Luis Obispo, California, United States
\item \Idef{org1329}Central China Normal University, Wuhan, China
\item \Idef{org14939}Centre de Calcul de l'IN2P3, Villeurbanne, France
\item \Idef{org1197}Centro de Aplicaciones Tecnol\'{o}gicas y Desarrollo Nuclear (CEADEN), Havana, Cuba
\item \Idef{org1242}Centro de Investigaciones Energ\'{e}ticas Medioambientales y Tecnol\'{o}gicas (CIEMAT), Madrid, Spain
\item \Idef{org1244}Centro de Investigaci\'{o}n y de Estudios Avanzados (CINVESTAV), Mexico City and M\'{e}rida, Mexico
\item \Idef{org1335}Centro Fermi -- Centro Studi e Ricerche e Museo Storico della Fisica ``Enrico Fermi'', Rome, Italy
\item \Idef{org17347}Chicago State University, Chicago, United States
\item \Idef{org1288}Commissariat \`{a} l'Energie Atomique, IRFU, Saclay, France
\item \Idef{org15782}COMSATS Institute of Information Technology (CIIT), Islamabad, Pakistan
\item \Idef{org1294}Departamento de F\'{\i}sica de Part\'{\i}culas and IGFAE, Universidad de Santiago de Compostela, Santiago de Compostela, Spain
\item \Idef{org1106}Department of Physics Aligarh Muslim University, Aligarh, India
\item \Idef{org1121}Department of Physics and Technology, University of Bergen, Bergen, Norway
\item \Idef{org1162}Department of Physics, Ohio State University, Columbus, Ohio, United States
\item \Idef{org1300}Department of Physics, Sejong University, Seoul, South Korea
\item \Idef{org1268}Department of Physics, University of Oslo, Oslo, Norway
\item \Idef{org1312}Dipartimento di Fisica dell'Universit\`{a} and Sezione INFN, Turin, Italy
\item \Idef{org1132}Dipartimento di Fisica dell'Universit\`{a} and Sezione INFN, Bologna, Italy
\item \Idef{org1145}Dipartimento di Fisica dell'Universit\`{a} and Sezione INFN, Cagliari, Italy
\item \Idef{org1315}Dipartimento di Fisica dell'Universit\`{a} and Sezione INFN, Trieste, Italy
\item \Idef{org1285}Dipartimento di Fisica dell'Universit\`{a} `La Sapienza' and Sezione INFN, Rome, Italy
\item \Idef{org1154}Dipartimento di Fisica e Astronomia dell'Universit\`{a} and Sezione INFN, Catania, Italy
\item \Idef{org1270}Dipartimento di Fisica e Astronomia dell'Universit\`{a} and Sezione INFN, Padova, Italy
\item \Idef{org1290}Dipartimento di Fisica `E.R.~Caianiello' dell'Universit\`{a} and Gruppo Collegato INFN, Salerno, Italy
\item \Idef{org1103}Dipartimento di Scienze e Innovazione Tecnologica dell'Universit\`{a} del Piemonte Orientale and Gruppo Collegato INFN, Alessandria, Italy
\item \Idef{org1114}Dipartimento Interateneo di Fisica `M.~Merlin' and Sezione INFN, Bari, Italy
\item \Idef{org1237}Division of Experimental High Energy Physics, University of Lund, Lund, Sweden
\item \Idef{org1192}European Organization for Nuclear Research (CERN), Geneva, Switzerland
\item \Idef{org1227}Fachhochschule K\"{o}ln, K\"{o}ln, Germany
\item \Idef{org1122}Faculty of Engineering, Bergen University College, Bergen, Norway
\item \Idef{org1136}Faculty of Mathematics, Physics and Informatics, Comenius University, Bratislava, Slovakia
\item \Idef{org1274}Faculty of Nuclear Sciences and Physical Engineering, Czech Technical University in Prague, Prague, Czech Republic
\item \Idef{org1229}Faculty of Science, P.J.~\v{S}af\'{a}rik University, Ko\v{s}ice, Slovakia
\item \Idef{org1184}Frankfurt Institute for Advanced Studies, Johann Wolfgang Goethe-Universit\"{a}t Frankfurt, Frankfurt, Germany
\item \Idef{org1215}Gangneung-Wonju National University, Gangneung, South Korea
\item \Idef{org1212}Helsinki Institute of Physics (HIP) and University of Jyv\"{a}skyl\"{a}, Jyv\"{a}skyl\"{a}, Finland
\item \Idef{org1203}Hiroshima University, Hiroshima, Japan
\item \Idef{org1254}Indian Institute of Technology Bombay (IIT), Mumbai, India
\item \Idef{org36378}Indian Institute of Technology Indore (IIT), Indore, India
\item \Idef{org1266}Institut de Physique Nucl\'{e}aire d'Orsay (IPNO), Universit\'{e} Paris-Sud, CNRS-IN2P3, Orsay, France
\item \Idef{org1277}Institute for High Energy Physics, Protvino, Russia
\item \Idef{org1249}Institute for Nuclear Research, Academy of Sciences, Moscow, Russia
\item \Idef{org1320}Nikhef, National Institute for Subatomic Physics and Institute for Subatomic Physics of Utrecht University, Utrecht, Netherlands
\item \Idef{org1250}Institute for Theoretical and Experimental Physics, Moscow, Russia
\item \Idef{org1230}Institute of Experimental Physics, Slovak Academy of Sciences, Ko\v{s}ice, Slovakia
\item \Idef{org1127}Institute of Physics, Bhubaneswar, India
\item \Idef{org1275}Institute of Physics, Academy of Sciences of the Czech Republic, Prague, Czech Republic
\item \Idef{org1139}Institute of Space Sciences (ISS), Bucharest, Romania
\item \Idef{org27399}Institut f\"{u}r Informatik, Johann Wolfgang Goethe-Universit\"{a}t Frankfurt, Frankfurt, Germany
\item \Idef{org1185}Institut f\"{u}r Kernphysik, Johann Wolfgang Goethe-Universit\"{a}t Frankfurt, Frankfurt, Germany
\item \Idef{org1177}Institut f\"{u}r Kernphysik, Technische Universit\"{a}t Darmstadt, Darmstadt, Germany
\item \Idef{org1256}Institut f\"{u}r Kernphysik, Westf\"{a}lische Wilhelms-Universit\"{a}t M\"{u}nster, M\"{u}nster, Germany
\item \Idef{org1246}Instituto de Ciencias Nucleares, Universidad Nacional Aut\'{o}noma de M\'{e}xico, Mexico City, Mexico
\item \Idef{org1247}Instituto de F\'{\i}sica, Universidad Nacional Aut\'{o}noma de M\'{e}xico, Mexico City, Mexico
\item \Idef{org23333}Institut of Theoretical Physics, University of Wroclaw
\item \Idef{org1308}Institut Pluridisciplinaire Hubert Curien (IPHC), Universit\'{e} de Strasbourg, CNRS-IN2P3, Strasbourg, France
\item \Idef{org1182}Joint Institute for Nuclear Research (JINR), Dubna, Russia
\item \Idef{org1143}KFKI Research Institute for Particle and Nuclear Physics, Hungarian Academy of Sciences, Budapest, Hungary
\item \Idef{org1199}Kirchhoff-Institut f\"{u}r Physik, Ruprecht-Karls-Universit\"{a}t Heidelberg, Heidelberg, Germany
\item \Idef{org20954}Korea Institute of Science and Technology Information, Daejeon, South Korea
\item \Idef{org1160}Laboratoire de Physique Corpusculaire (LPC), Clermont Universit\'{e}, Universit\'{e} Blaise Pascal, CNRS--IN2P3, Clermont-Ferrand, France
\item \Idef{org1194}Laboratoire de Physique Subatomique et de Cosmologie (LPSC), Universit\'{e} Joseph Fourier, CNRS-IN2P3, Institut Polytechnique de Grenoble, Grenoble, France
\item \Idef{org1187}Laboratori Nazionali di Frascati, INFN, Frascati, Italy
\item \Idef{org1232}Laboratori Nazionali di Legnaro, INFN, Legnaro, Italy
\item \Idef{org1125}Lawrence Berkeley National Laboratory, Berkeley, California, United States
\item \Idef{org1234}Lawrence Livermore National Laboratory, Livermore, California, United States
\item \Idef{org1251}Moscow Engineering Physics Institute, Moscow, Russia
\item \Idef{org1322}National Centre for Nuclear Studies, Warsaw, Poland
\item \Idef{org1140}National Institute for Physics and Nuclear Engineering, Bucharest, Romania
\item \Idef{org1165}Niels Bohr Institute, University of Copenhagen, Copenhagen, Denmark
\item \Idef{org1109}Nikhef, National Institute for Subatomic Physics, Amsterdam, Netherlands
\item \Idef{org1283}Nuclear Physics Institute, Academy of Sciences of the Czech Republic, \v{R}e\v{z} u Prahy, Czech Republic
\item \Idef{org1264}Oak Ridge National Laboratory, Oak Ridge, Tennessee, United States
\item \Idef{org1189}Petersburg Nuclear Physics Institute, Gatchina, Russia
\item \Idef{org1170}Physics Department, Creighton University, Omaha, Nebraska, United States
\item \Idef{org1157}Physics Department, Panjab University, Chandigarh, India
\item \Idef{org1112}Physics Department, University of Athens, Athens, Greece
\item \Idef{org1152}Physics Department, University of Cape Town and  iThemba LABS, National Research Foundation, Somerset West, South Africa
\item \Idef{org1209}Physics Department, University of Jammu, Jammu, India
\item \Idef{org1207}Physics Department, University of Rajasthan, Jaipur, India
\item \Idef{org1200}Physikalisches Institut, Ruprecht-Karls-Universit\"{a}t Heidelberg, Heidelberg, Germany
\item \Idef{org1325}Purdue University, West Lafayette, Indiana, United States
\item \Idef{org1281}Pusan National University, Pusan, South Korea
\item \Idef{org1176}Research Division and ExtreMe Matter Institute EMMI, GSI Helmholtzzentrum f\"ur Schwerionenforschung, Darmstadt, Germany
\item \Idef{org1334}Rudjer Bo\v{s}kovi\'{c} Institute, Zagreb, Croatia
\item \Idef{org1298}Russian Federal Nuclear Center (VNIIEF), Sarov, Russia
\item \Idef{org1252}Russian Research Centre Kurchatov Institute, Moscow, Russia
\item \Idef{org1224}Saha Institute of Nuclear Physics, Kolkata, India
\item \Idef{org1130}School of Physics and Astronomy, University of Birmingham, Birmingham, United Kingdom
\item \Idef{org1338}Secci\'{o}n F\'{\i}sica, Departamento de Ciencias, Pontificia Universidad Cat\'{o}lica del Per\'{u}, Lima, Peru
\item \Idef{org1271}Sezione INFN, Padova, Italy
\item \Idef{org1133}Sezione INFN, Bologna, Italy
\item \Idef{org1286}Sezione INFN, Rome, Italy
\item \Idef{org1115}Sezione INFN, Bari, Italy
\item \Idef{org1313}Sezione INFN, Turin, Italy
\item \Idef{org1316}Sezione INFN, Trieste, Italy
\item \Idef{org1155}Sezione INFN, Catania, Italy
\item \Idef{org1146}Sezione INFN, Cagliari, Italy
\item \Idef{org36377}Nuclear Physics Group, STFC Daresbury Laboratory, Daresbury, United Kingdom
\item \Idef{org1258}SUBATECH, Ecole des Mines de Nantes, Universit\'{e} de Nantes, CNRS-IN2P3, Nantes, France
\item \Idef{org1304}Technical University of Split FESB, Split, Croatia
\item \Idef{org1168}The Henryk Niewodniczanski Institute of Nuclear Physics, Polish Academy of Sciences, Cracow, Poland
\item \Idef{org17361}The University of Texas at Austin, Physics Department, Austin, TX, United States
\item \Idef{org1173}Universidad Aut\'{o}noma de Sinaloa, Culiac\'{a}n, Mexico
\item \Idef{org1296}Universidade de S\~{a}o Paulo (USP), S\~{a}o Paulo, Brazil
\item \Idef{org1149}Universidade Estadual de Campinas (UNICAMP), Campinas, Brazil
\item \Idef{org1239}Universit\'{e} de Lyon, Universit\'{e} Lyon 1, CNRS/IN2P3, IPN-Lyon, Villeurbanne, France
\item \Idef{org1205}University of Houston, Houston, Texas, United States
\item \Idef{org20371}University of Technology and Austrian Academy of Sciences, Vienna, Austria
\item \Idef{org1222}University of Tennessee, Knoxville, Tennessee, United States
\item \Idef{org1310}University of Tokyo, Tokyo, Japan
\item \Idef{org1318}University of Tsukuba, Tsukuba, Japan
\item \Idef{org21360}Eberhard Karls Universit\"{a}t T\"{u}bingen, T\"{u}bingen, Germany
\item \Idef{org1225}Variable Energy Cyclotron Centre, Kolkata, India
\item \Idef{org1306}V.~Fock Institute for Physics, St. Petersburg State University, St. Petersburg, Russia
\item \Idef{org1323}Warsaw University of Technology, Warsaw, Poland
\item \Idef{org1179}Wayne State University, Detroit, Michigan, United States
\item \Idef{org1260}Yale University, New Haven, Connecticut, United States
\item \Idef{org1332}Yerevan Physics Institute, Yerevan, Armenia
\item \Idef{org15649}Yildiz Technical University, Istanbul, Turkey
\item \Idef{org1301}Yonsei University, Seoul, South Korea
\item \Idef{org1327}Zentrum f\"{u}r Technologietransfer und Telekommunikation (ZTT), Fachhochschule Worms, Worms, Germany
\end{Authlist}
\endgroup

%
%
\end{document}